\documentstyle[12pt]{article}
\begin{document}
\begin{titlepage}

\center{Molecular dynamics simulation
for  pressure-induced structural
transition from C$_{60}$ fullerene into amorphous diamond}

\center{Akihito Kikuchi and Shinji Tsuneyuki}
\center{Institute for Solid State Physics, University of Tokyo,}
\center{Kashiwa-no-ha  5-1-5, Kashiwa-shi, Chiba 277-8581, Japan}

\normalsize
\abstract{
The pressure-induced structural transition in fcc C$_{60}$ fullerene 
by shock compression and rapid quenching
is investigated by a semi-empirical
tight-binding molecular dynamics simulation,
adopting a constant-pressure scheme and a method of the order N
electronic structure calculation.
At first, the process of the amorphization
of C$_{60}$  is demonstrated.
The simulated results indicated that,
in the material fabricated after the quenching,
the remaining dangling bonds
have a large influence on physical properties,
such as, the density and the presence of the
band gap at the Fermi level.
We have furthermore studied the formation 
of the short-range order, observed as amorphous diamond.
In order to form the amorphous diamond phase,
the bonding state of sp$^2$ must be turned into 
that of sp$^3$. The transition process is
seriously influenced from the 
the external pressure, the temperature,
or the presence of hydrogen.
The comparison to the pressure-induced structural transition
in the graphite is also executed and a brief discussion
on the difference  in those carbon crystals is given. 
}
\end{titlepage}

\section{Introduction}
\setlength{\baselineskip}{1.0cm}
In the field of high-pressure material science, 
diverse carbon systems under pressure
have been intensively studied 
with interest in synthesizing new phases.
For example,
the pressure-induced structural transition from
C$_{60}$ fullerene to amorphous diamond 
is realized by  shock compression and rapid quenching\cite{HIRAI}. 
The shock compression and rapid quenching 
generate the high pressure (50-55GPa) and the temperature
(2000-3000K) in a fraction of a microsecond. 
Consequently the C$_{60}$ fcc crystal transforms
into  stable transparent glassy chips of $\mu m$ in size,  
which are confirmed to be
an amorphous phase of diamond by electron energy-loss spectroscopy 
and electron diffractometry\cite{HIRAI}. 
Though this phase has a short-range order similar to  normal diamond,
it is amorphous.
This amorphous phase is characterized by
being consisted mostly from sp$^3$ bonding,
in contrast with  previously reported amorphous
carbon which is considered to be a
disordered phase of graphite with sp$^2$ bonds\cite{GRA_AM,GRA_AM2}.
By an investigation of the sample after the shock compression,
the fabrication process for the amorphous diamond
is postulated in the following way \cite{HIRAI}.
Once the fcc cell of  C$_{60}$ is compressed,
the inter-cluster bonding between $C_{60}$ molecules is formed. 
In other words, the polymerization of the
C$_{60}$ clusters occurs there,
as was observed 
in several high-pressure experiments\cite{POLYM,POLYM1,POLYM2}.
By the further compression,
molecules collapse and change into the amorphous.
In the amorphous, the bonding state of sp$^3$  is gradually produced
and it forms a short-range order extending up to
the volume of the normal diamond unit cell.
This high temperature phase
is quenched and  obtained as the amorphous diamond,
in which 
a greater part of the bonding state turns into sp$^3$ type
without the crystal growth in long range.

In fact, there is uncertainty in the understanding of
the fabrication process from C$_{60}$ to amorphous diamond,
since the postulation for the process
is obtained by checking the various phases 
of C$_{60}$ remaining after the compression and the quenching
and it is not based on direct observation of the reaction.
In order to clarify such a transition process, 
the aid of the computer simulation will be needed.
As for the structural transformation in carbon systems,
there are a lot of theoretical approaches, 
from the first principles or semi-empirical way.
However, those studies focus on the
transformation between graphite and diamond. 
In the present stage, the theoretical simulation
on the pressure-induced structural transformation in
C$_{60}$ has not been executed sufficiently.
Therefore, the object of the present work is set to be a theoretical
understanding of the pressure-induced structural
transition in C$_{60}$ crystal.
For this purpose, a molecular dynamics simulation is  executed
using a tight-binding Hamiltonian with model parameters for carbon 
and hydrogen systems 
proposed by Winn , Rassinger and Hafner\cite{PARAM_CH},
combined with the constant pressure scheme 
by Wentzcovotch\cite{Wentz,G2D}.
The model parameters can reproduce the 
energy difference and geometries 
in various carbon systems including hydrogen,
such as molecules and reconstructed diamond surfaces,
as well as the bulk property of diamonds and graphites.
This model parameters also give  results consistent with
other ab-initio and semi-empirical calculations for liquid carbon phase.
In addition, to save computational costs, 
a part of calculations, involved in the 
crystal structure optimization,
was carried out using the density matrix method \cite{DENS_MAT},
what is called the order N method.

The contents of this paper are as follows.
At first the process of 
the pressure-induced amorphization of C$_{60}$ will be presented.
According to the simulation presented here,
the C$_{60}$ fcc crystal  is easily and speedily
transformed into the amorphous by the compression,
going through the path postulated by the experiment as above,
since the  C$_{60}$ structure is constructed by bondings of 
distorted sp$^2$ type and it has potentiality
in the transition to sp$^3$ type bondings.
However, in the simulated result, the amorphous phase itself
is far from the state which should be classified as "amorphous diamond",
especially just after the transition from C$_{60}$,
since the collapsed C$_{60}$ fcc crystal 
includes  a number of dangling bonds which have not 
yet turned into  sp$^3$ bondings.
Therefore, the present work will furthermore pursue
the transition from sp$^2$ to sp$^3$ in the amorphous phase 
through the compression,
which will gradually reduce the number of the 
dangling bonds and drive the system towards
the "amorphous diamond" phase of randomly packing
sp$^3$ bondings. The details of those
dynamical effects, such as the dependence on the temperature and the
pressure, are not necessarily clarified by the experiment alone.
Finally, the present paper will give a comparison
of the structural transition in C$_{60}$ to 
the pressure-induced graphite-diamond transition by the simulation.
It will explain the reason why  
 C$_{60}$ fcc crystal shows the transition to  
amorphous diamond phase and why graphite directly turns
into perfect cubic diamond.

\section{The simulation scheme}
In the present work, the simulations are
executed in the following way.

1)The initial external pressure is  set to be 0GPa
and the pressure is gradually raised to the desired value.

2) Then the simulation is executed at a stationary pressure
in  sufficiently long time.

3) To simulate the process of the rapid quenching,
the optimized structure at 0GPa and 0K is finally obtained,
by letting down the applied pressure.

In the present stage, we do not have the detailed information on the 
physical quantities in the sample
through the shock-compression, 
such as, the variation in the pressure and the temperature
related to the propagation of shock waves. 
For example, the temperature is not preserved
and rapidly escapes from  the compressed system
while the  pressure is kept to be high.
Thus, in some calculations,
the temperature is also given as a parameter independent of
the pressure.
In the electronic structure calculation,
a periodic boundary condition is applied to the 
fcc unit cell containing four C$_{60}$ molecules and the summation 
over k-space is estimated at the $\Gamma$ point alone.

\section{Amorphization from the C$_{60}$ fcc crystal}
\setlength{\baselineskip}{1.0cm}
\subsection{ Compression at 0K}
At first,
the  C$_{60}$ fcc crystal was optimized at the pressure of 50-100GPa,
keeping the temperature at 0K. 
The C$_{60}$ fcc  structure 
is stable in this pressure range. 
The initial cubic cell parameter including
four C$_{60}$ molecules is set to be 14.15\AA.
For example, a compression at 50GPa   
changed the cell into smaller one, in which
the cell parameter shrinks to 13.6 \AA.
In the compressed cell, the relative positions of atoms
are still preserved, while the covalent bonds are formed between 
C$_{60}$ molecules.
This is a kind of the polymerization, but not the amorphization.

\subsection{Compression at finite temperature}
Obviously, the reason why the C$_{60}$ crystal
is stable in such a high pressure 
is that the simulation is executed in an 
unphysical condition, i.e.,  keeping the temperature at 0 K.
In the experiments for fabricating the amorphous diamond,
the temperature rapidly rises from the room temperature
to the range no less than 2000-3000 K.
Then the collapse of the crystal will become drastic
owing to the active movements of atoms,
as can be expected from the phase diagram of carbon\cite{Bundy}. 
Therefore, in this section, the simulation is executed
with the condition that the temperature of the system can vary. 
To do this, the equation of motion is solved in the 
constant pressure scheme of molecular dynamics\cite{Wentz}
without any scaling of the kinetic energy.

Figure 1 shows the snapshots of 
an example of the amorphization.
In this case, the pressure is linearly increased from 0GPa
to 65GPa in the first 0.45ps and kept constant after that.
The temperature is  set to be 300K in the beginning and 
spontaneously increases to 2000-3000K by compression
partly because this is adiabatic compression
and partly because of recombination of C-C bonds.
In the early stage,
the stacking structure of the fcc crystal is still preserved. 
The C$_{60}$ molecules come to show  deformation 
and  polymerization,
if the inter-cluster distance
becomes much closer to the intra-cluster atomic distance,
as was observed in the compression study by Duclos {\it et al.}\cite{DUCLOS}.
By  further compression, 
 amorphization occurs and it results in the 
destruction of  the molecular structure in the whole cell.
At the same time, 
the sp$^2$ bonding gradually transforms into the sp$^3$. 

The structural change in Fig.1 is also quantitatively checked.
Figures 2 and 3  show the
change in the pair-distribution function
and the bond-angle distribution,
which correspond to the snapshot of the 
transition process in Fig.1.
The initial fcc C$_{60}$ crystal has already
turned into the amorphous phase in the first 1000 MD steps (1.0ps),
as can be seen from these figures.
The initial bond-angle distribution  has two peaks
at 108 degrees and 120 degrees, which mean
bond-angles in five- and six-membered rings. 
As the time goes by, these two clear peaks
become more  and more broad and finally merge
into one peak around 110 degrees. This indicates the
vanishing of the molecular structure of C$_{60}$.

Compressions in the pressure higher than 65GPa
accompanies more temperature increase 
and it turns C$_{60}$  into  amorphous.
On the other hand, if  the maximum pressures are
lower than  50GPa, and 
if the simulation starts at the room temperature, 
the fcc C$_{60}$ crystal does
not collapse completely,
probably because the final temperature(1000-2000K) is not sufficiently high.
However, if the initial temperature is set to be sufficiently high,
as high as  in the order of 1000K,
the amorphization proceeds even at about 50GPa.

\section{Formation of amorphous diamond}
To gain the frozen phase after the rapid-quenching,
we execute the crystal structure optimization
firstly at the high pressure   
and finally obtain the optimized structure at 0GPa,
gradually letting down the pressure.
In the actual process of the
shock compression and rapid quenching, the high-temperature
phase is frozen into the system owing to the
rapid decrease in the temperature, while the
high pressure is still kept.

Figures 4(a) and 4(b) show the results of the crystal structure
optimization,
where the maximum compression pressures
are set to be 65GPa and 125GPa, respectively.
(In both cases, the optimized structures
at 0GPa are obtained after the MD simulations
continued as long as  about 10ps,
so that the property of the finally fabricated material
shall not seriously be affected by the finiteness
of the simulation time.)
In these structures,  
the short-range order, formed in the high temperature
and the high pressure,
remains almost unchanged.
The pair distribution function (not shown here)
shows the peaks around 1.5\AA { }   and 2.5\AA, which
correspond to the contributions from
the first- and next-nearest neighboring atoms,
as is seen in that of the high temperature phase.
However, only from the location of the peaks,
it is difficult to  determine
whether the short-range order comes from a diamond-like structure
or not. It may be likely that the
peaks are attributed to the fragments of the C$_{60}$,
since the distances
from one atom to the first- or next-nearest neighboring atoms
are of the same extent both in 
diamond and in a single C$_{60}$ molecule.
Thus the distribution of the azimuthal angle,
as is defined in Fig.5(a),  was checked.
Here the azimuthal angle is defined to be 
the relative angle of the two planes,
respectively spanned by two bonds, when these two plane
share one common bond. 
This distribution gives
the information on the networking structure of the
tetrahedron of sp$^3$, and it will have   peaks 
around  60  degrees or 180 degrees for cubic diamond and 
around 140 degrees for  the C$_{60}$ structure.
(The pairs of six- and five-membered rings in C$_{60}$ form the
angle of about 140 degrees.)
In  Fig.5(b), for the case of the compression at 65GPa,
there are no prominent peaks around 60 or 140 degrees.
It means that
the cage structure of C$_{60}$  vanishes, while
the diamond-like short range order is not clearly formed.
On the other hand,  in the structure compressed at 125GPa, 
there is a prominent peak near 60 degrees in Fig.5(c).
The cubic-diamond-like short-range order, extending up to the 
third-nearest neighbors,
is much more developed there,
since this structure compressed by the larger pressure
has much more number of sp$^3$ bonding.

In fact, the networking structure of bonds 
is not necessarily confined to cubic diamond phase alone,
since the energy difference between cubic and hexagonal diamond phase
is little \cite{G2D}. Several local bonding structures 
similar to that of hexagonal diamond are found in the simulated results.
In perfect hexagonal diamond structure,
the azimuthal angle with respect to a bond on an atom 
parallel to the c-axis takes 0 and 120 degrees.
On the other hand 
azimuthal angles with respect to other three bonds on an atom 
take 60 and 180 degrees, which is the same as cubic diamond.
Therefore, even in perfect hexagonal diamond phase, the
contributions to the distribution of azimuthal angles
at 0 and 120 degrees is weak and 
it is  one third of those at 60 and 180 degrees.
Thus contribution from hexagonal-diamond like short range 
order around 0 or 120 degrees is hidden 
due to the randomness in amorphous.

The density in the structure compressed
by the maximum pressure 65GPa [Fig.4(a)] 
is estimated to be  about 2.7 g/ cm$^3$. 
On  the  other hand,
the density is 3.5 g/ cm$^3$ 
after the compression at the maximum pressure 125GPa [Fig.4(b)],
which is comparable to the experiment. 
In the actual amorphous diamond,
the density is estimated to be larger than 3.3 g/cm$^3$\cite{KAGAKU}.

Figures 6(a) and 6(b) show the density of states 
of the finally fabricated material (frozen phase)
after the compression at the maximum pressure 65GPa and 125GPa.
These electronic structures
are quantitatively different from
that of the initial C$_{60}$ fcc crystal,
which is  given in Ref.\cite{C60FCCBAND}.
The difference between Fig.6(a) and 6(b) results from the
number of dangling bonds.
In these figures, the contribution to the DOS
from threefold carbon atoms 
is compared to the total DOS.
Owing to the contribution from dangling bonds,
the gap between the conduction and valence bands vanishes.
In case of the compression at 65GPa[Fig.6(a)],
the ratio of atoms with dangling bond amounts to 50 \%,
while in case of the compression at 125GPa[Fig.6(b)],
the ratio of such atoms deceases to about 10\%.
Therefore, in the latter case, the conduction and valence
bands are distinguished by a reduction of the DOS
near the Fermi level, which is featured as a shaded zone in the figure.
This reduction in the DOS is interpreted to be 
an analogous of the wide gap in the perfect diamond structure.
In Fig.6(a), a narrow gap at the Fermi level
is formed in the dangling bond states
and it separates the occupied and unoccupied states.
This is because the presence of a gap, even if it is  narrow,
makes the system more stable.

Comparing the results corresponding to the compression
at 65GPa and 125GPa,
the latter case,  compressed at the larger pressure,
is considered to be a better simulation,
since the actual
amorphous diamond has a transparent optical property
similar to the normal diamond
and it will have an electronic structure
with a large gap near the Fermi level. 
In our simulations, 
at  the pressure above 100GPa  and the temperature above 5000K,
the ratio of threefold carbon atoms 
decreases to 10$-$20\%, and 
the valence and conduction bands tend to be distinguished 
by a wide range reduction in the DOS near the Fermi level,
owing to the decrease of the dangling bonds.
In the experimentally obtained amorphous diamond,
the number of the the dangling bonds will also be 
reduced, possibly more than in this simulation.
(It is certain that the dangling bonds still remain
in the actual amorphous diamond, as was indicated by
the EELS spectrum in  Ref.\cite{HIRAI}.)

\section{An analysis for 
the dynamical process in the transition}
After the rapid amorphous transition,
it takes a long time until the short range order is formed again.
In this section, the dynamical process in the reaction,
especially related to  the pressure and the temperature, is investigated. 
Figure 7 shows the results by a simulation where the
external pressure is increased from 0GPa to 65GPa,
and after that, the pressure is furthermore increased to 125 GPa, 
in order to check the pressure dependence. 
Figures 7(a)-(d) show the time dependence of 
the external pressure, the temperature,
the mean square of the displacement, and 
the ratio of carbon atoms with fourfold coordination.
The increase in the ratio
of carbon atoms with fourfold coordination
stands for the transition from sp$^2$ to  sp$^3$,
and it reflects on the reconstruction of the short-range order.
In the amorphous phase, sp$^2$-type bonding is gradually  
transformed into sp$^3$-type, while the transition speed,
in other word, the speed of the short-range order formation
is going down. 
However, at least in the order of the picoseconds,
the system has not yet arrive at the stationary state.
The figure furthermore shows 
the dependence of the reaction on the external pressure.
When the external pressure is raised again, 
the number of the sp$^3$ bonding grows much more,
since the transition is accelerated by the
higher temperature and the higher density.
The density in the high temperature phase
amounts to about 3.0g/cm$^3$ at 65GPa
and increases to about 3.5g/cm$^3$ at 125GPa.

The simulated result supports the postulations
for the amorphous diamond formation process
from the experimental data\cite{HIRAI}.
We should stress here the importance of the following
phenomena. Since the high temperature is generated through
the reconstruction of the bonding, 
the reaction is enhanced so that
the system can cross the potential barrier and transform its structure.
(The system obtains work by the compression and the
temperature increases to some extent, but it is not enough to speed 
up the reaction furthermore.)
As an example, 
figure 8 shows the results of the
two simulation  where the external pressure
is increased to  125GPa. 
In the first case, denoted as (A), 
there is no restriction to the variation of the temperature.
On the other hand , in the second case,  denoted as (B),
the temperature is set to be 2500K after 0.5ps.
The transition from sp$^2$  to sp$^3$ is
apparently hindered in the case of (B),
since the movement of atoms is inactive
because of the lower temperature, in contrast with the case (A). 
For the same reason,
the compression at 0K, shown in the previous section,
cannot turn C$_{60}$ to the amorphous.
These simulations also suggest that 
there are possibly innumerable quasi-stable configurations
by which the system is easily trapped.
The amorphous diamond phase can also 
be regarded as one of such quasi-stable transient phases
located in the reaction path from C$_{60}$ to
the bonding state of sp$^3$ in whole crystal,
i.e.  perfect diamond phase. 
(In fact, the shock compression  applied to 
lower-grade C$_{60}$ \cite{C60_DIA} exhibited the
entire transition to the diamond crystallite,
probably because of the easier crystal growth in the
presence of defects and impurities.)
Since shock compression process continues as long as  nanoseconds,
it is possible that the actual amorphous diamond phase
has a structure much closer to perfect diamond,
compared to the present simulation whose time-scale
is at most picoseconds order. 
In the actual shock compression,
a large fluctuation in the pressure and the temperature
through the shock-wave propagation 
will fabricate  variously altered phases,
as are classified in several states\cite{HIRAI},
ranging from the slightly compressed fcc structure 
to the amorphous diamond.

It appears that the speed of 
increasing pressure may have some influence on the
reaction.  For example, if the response of the crystal inner stress
cannot catch up with the rapid 
increase of the pressure, the decaying process of the crystal 
will become more drastic. 
The response of the inner stress is related to the cell deformation
and it is dependent on the fictitious mass assigned to the cell deformation, 
as well as the external pressure and the inner stress in the
constant-pressure scheme.
We would like to avoid such dependence of the simulation
on the artificial degree of freedom as possible.
For this purpose,
we adjusted the fictitious mass heavy enough so that  
the fluctuation of the "kinetic energy" assigned to the cell deformation
is kept to be very small  and the almost "isenthalpic"
simulation becomes possible at a stationary pressure. 
The speed of raising the applied pressure 
is set to be sufficiently slow
in such a way that the inner stress 
shall rise parallel with the external pressure. 
By doing so, the system property is almost determined by the
the current pressure, and scarcely dependent on 
the kinetic contribution from the cell deformation.
For example, 
in  figure 7 and 8, where
the pressure is raised to 125GPa in two different ways,
the final system properties, such as the temperature
and the bonding order, are similar.

\section{Discussions}

In the electronic structure of the finally fabricated material after
the compression,
the contribution from dangling bonds is not negligible.
If the contribution of this kind
appearing between the valence and conduction band
is large, the fabricated material will
lose the transparent optical property and it will not be qualified
to be called amorphous diamond.
As we have seen, the contribution from dangling bonds
will be reduced after the compression
at sufficiently higher temperature and pressure.
In fact, there may be other mechanism
that will reduce the contribution from dangling bonds.
For example, it is well-known that the presence of hydrogen atoms
reduces the number of dangling bonds in case of amorphous silicon.
The DOS after the shock-compression
in presence of hydrogen atoms are  given in Fig.9\cite{TEMP_CH}. 
A comparison between Fig.9(a) and Fig.6(a),
which are the DOS after the compression at 55GPa and 65GPa,
shows that the DOS is apparently reduced around the Fermi level 
in the presence  of the hydrogen, even if the pressure
is somewhat lower. 
However, there is no clear gap between the valence and conduction
band in Fig.9(a).
Such a situation is not improved so much
at higher pressure, as in Fig.9(b).
In addition, according to the simulation like this,
even in the case when  much more numbers 
of the hydrogen atoms are included, 
dangling bonds not terminated by hydrogens are still left.
This will be because the reactivity and the mobility
of bonds are weakened by the presence of
many hydrogen atoms and the formation
of the short range order of sp$^3$ will rather be hindered.

In case of the compression of C$_{60}$,
there are number of quasi-stable configurations in amorphous phase, 
and such a transient phase will easily 
be frozen by the quenching.
This will be the major difference to the pressure-induced structural
transition from graphite to diamond.
Figure 10 shows the transition of the bonding state
from sp$^2$ to sp$^3$ in the compression of the graphite.
In the figure, the rapid increase in the 
ratio of fourfold carbon atoms after 1000fs
stands for the transition from graphite to  diamond.
The transition speed is far faster than
the cooling speed of the temperature
in the shock compression experiment which is
estimated to be from 10$^6$ to 10$^{10}$ K/s\cite{QUENCH_SPEED}.  
Graphite and diamond phases are located in very close
configurations in the potential surface and there is 
no stable transient state in the transition path from 
graphite to diamond.
Since graphite rapidly transforms into the cubic diamond
without being trapped by any quasi-stable structure ,
perfect diamond phase remains alone after the quenching
and amorphous diamond phase will not be
obtained\cite{COMPRESS_GRA}.

It should be noted here that, 
in  comparison with ab-initio theory,
there are less accuracy and less transferability
in the tight-binding models,
in spite of the fact that the model parameters  succeed in
reproducing the bulk property of certain carbon crystals.
This is because the tight-binding model parameters
are obtained by the fitting to the ab-initio results
for several crystal structures.
In order to  check
the model parameter dependence of our simulation,
we have executed
additional calculations using another model parameter 
by Xu {\rm et al.} \cite{Xu} and 
compared the results to those given in
the previous sections.
According to the results obtained by the Xu's parameters,
if the initial condition and the applied pressure are the same, 
the electronic and structural properties,
such as the transition rate from sp$^2$ to sp$^3$
and the increase in the temperature, 
are of the same extent, compared to those given in the previous sections.
For example, when the compression is executed above 100GPa, 
the ratio of  atoms with dangling bond decreases to $10-20$ \%
and the DOS shows the reduction near the Fermi level,
as well as the result in the previous section.

It should also be mentioned that
the model Hamiltonian used here 
does not include van der Waals interactions and 
is not quantitatively sufficient for 
the initial C$_{60}$ fcc structure formed by
van der Waals force.
However, the purpose of the present work
is the investigation in a compressed carbon system.
Since the van der Waals force is much weaker
than covalent bonds between atoms,
it will be negligible in the simulation, especially 
after the transition into  the high-density amorphous.
In other words, the model parameters adopted here
will be a good, even if  not the best, description
for the physical process of the pressure-induced
structural transition from
the C$_{60}$ fcc crystal into the amorphous diamond.

\section{Conclusion}
We have investigated 
the amorphization of C$_{60}$ fcc 
and the formation of amorphous diamond phase.
The electronic property of
amorphized phase of C$_{60}$ just after the transition is
far from so-called amorphous diamond, since
the amorphized C$_{60}$ contains a number of dangling bonds
which have not yet turned into sp$^3$ type bonding.
Amorphous diamond phase is being gradually formed
through the change in the bonding structure from sp$^2$ to
sp$^3$ under high pressure and high temperature.
If dangling bonds are sufficiently reduced,
 the valence and conduction 
bands are distinguished and a transparent optical property 
is observed in amorphous diamond phase as well as perfect cubic diamond.
Amorphous phase is interpreted as one of 
quasi-stable phases in the transition path from C$_{60}$ to diamond
phase under pressure.
If the temperature and the pressure are not sufficiently high,
the reaction will be interrupted before forming amorphous diamond phase,
probably because there are many quasi-stable 
phases between C$_{60}$ and  diamond and 
the system is liable to be trapped by such phases.
This is in contrast with pressure-induced graphite-diamond transition.
Since graphite and diamond phase are located in very near configurations
in the potential surface,  the transition from graphite
to diamond proceeds rapidly without passing through any stable transient phase.
From this reason, the formation of amorphous diamond phase
is characteristic in the compression of C$_{60}$ and
such a phase is not obtained  in the compression of graphite.

\newpage
\section*{Figure captions}
\setlength{\baselineskip}{14.5pt}
\begin{itemize}
\item Figure 1: Snapshots of the transition from
                C$_{60}$ fcc crystal  to amorphous are given.
                The pressure is increased from 0GPa to 65GPa in 0.45ps
                and kept to be 65GPa after that.           
                The pictures show the system cut out into
                a $15\times 15\times 15$\AA$^3$ cube,
                which does not stand for the unit cell.

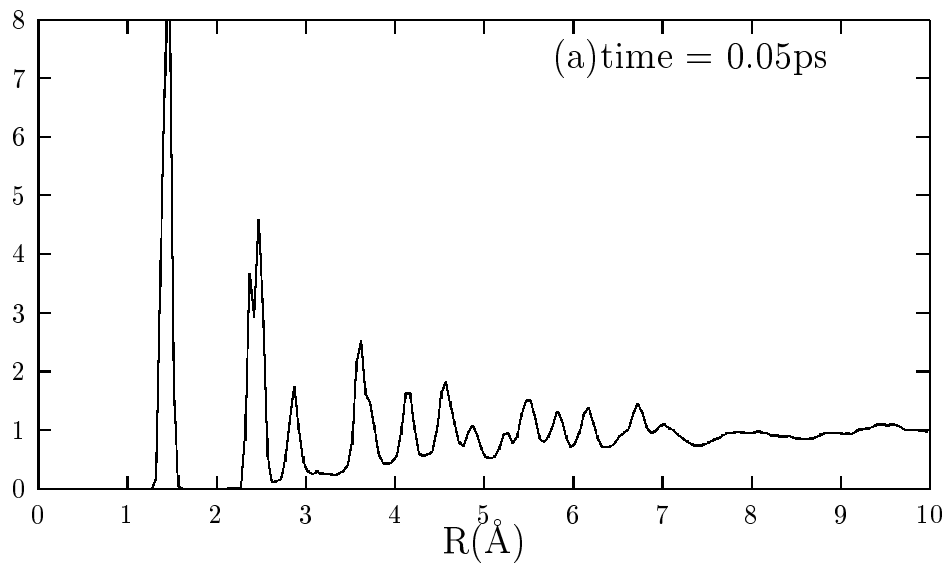
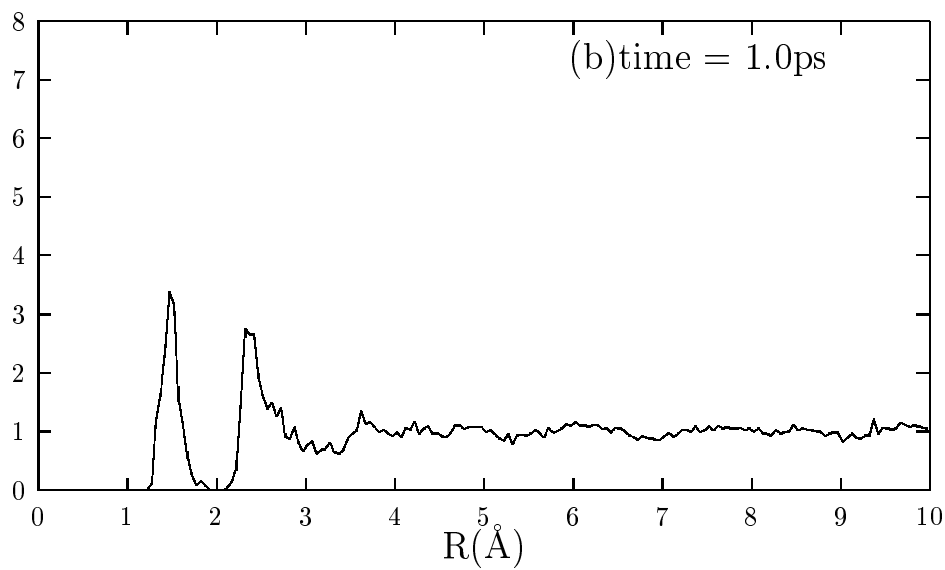
\item Figure 2 : Pair distribution function in each MD step,
                 which  corresponds to the compression
                 given in figure 1. 

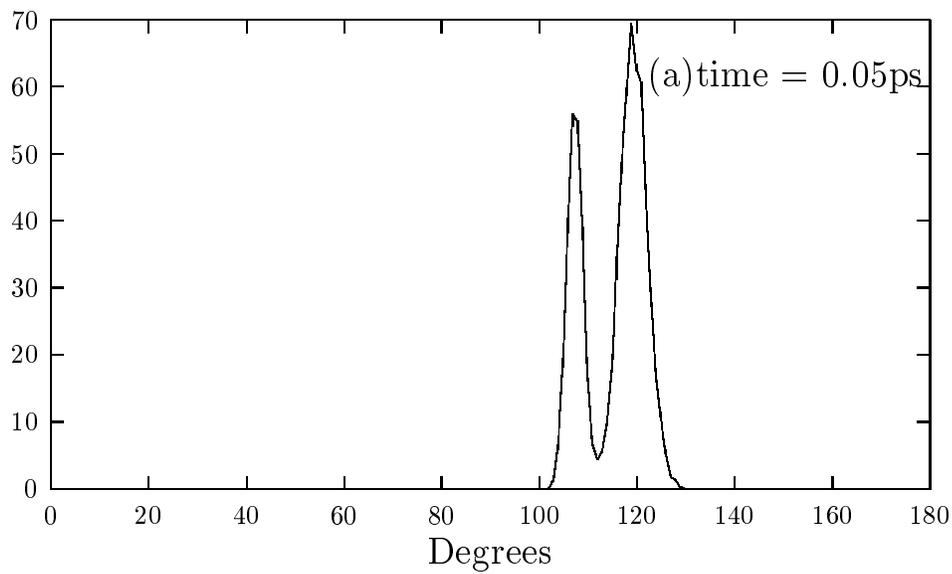
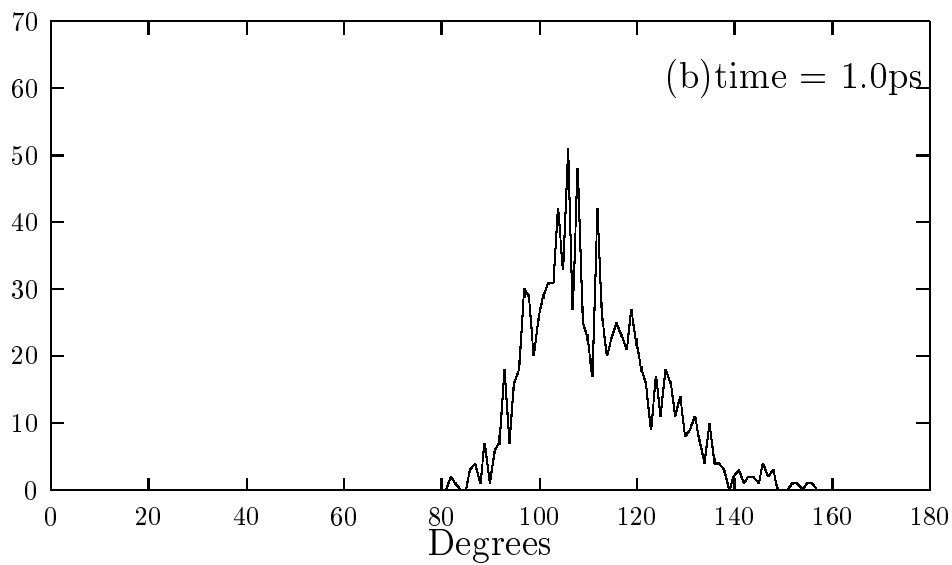
\item Figure 3: Bond-angle distribution in each MD step,
                which correspond to the compression
                given in figure 1.
                Although the structures, characteristic to C$_{60}$,
                remain at first, they have almost vanished
                in 0.5ps and the system turns into the amorphous.

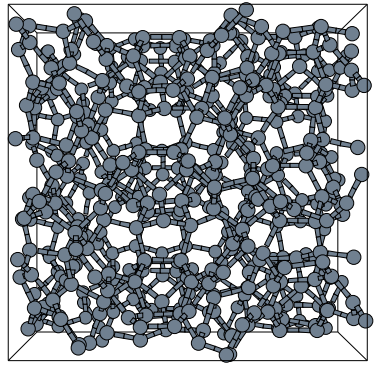
\item Figure 4(a): The structure fabricated after the compression
                   at the maximum pressure at 65GPa.

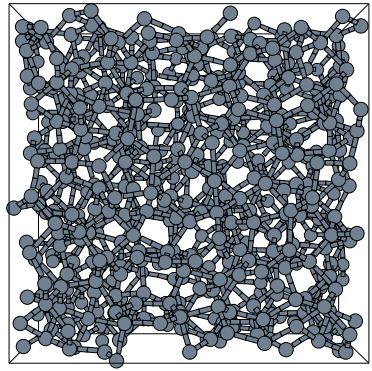
\item Figure 4(b): The structure fabricated after the compression
                   at the maximum pressure at 125GPa.

\item Figure 5(a):Definition of azimuthal angle. 

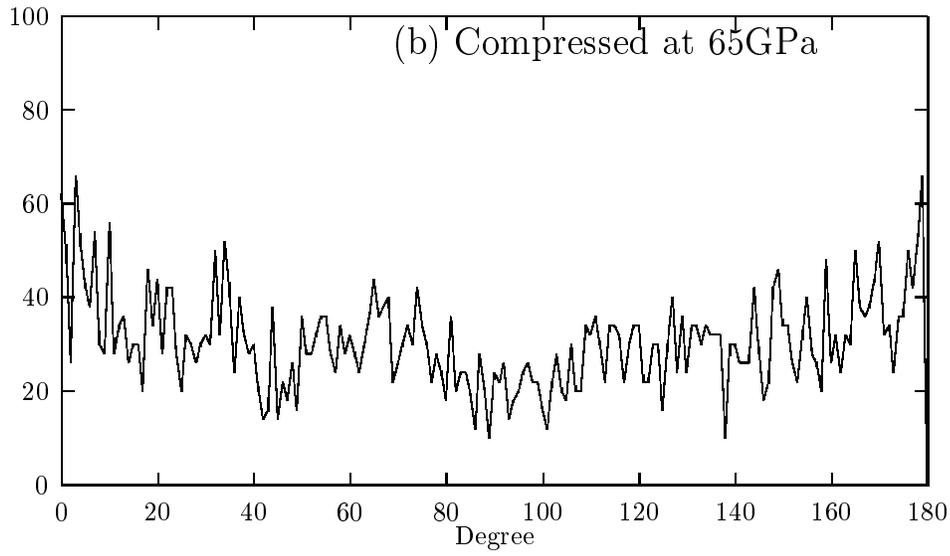
\item Figure 5(b):Distribution of the azimuthal angle 
                 after the compression at 65GPa, 
                 in the structure of figure 4(a). 

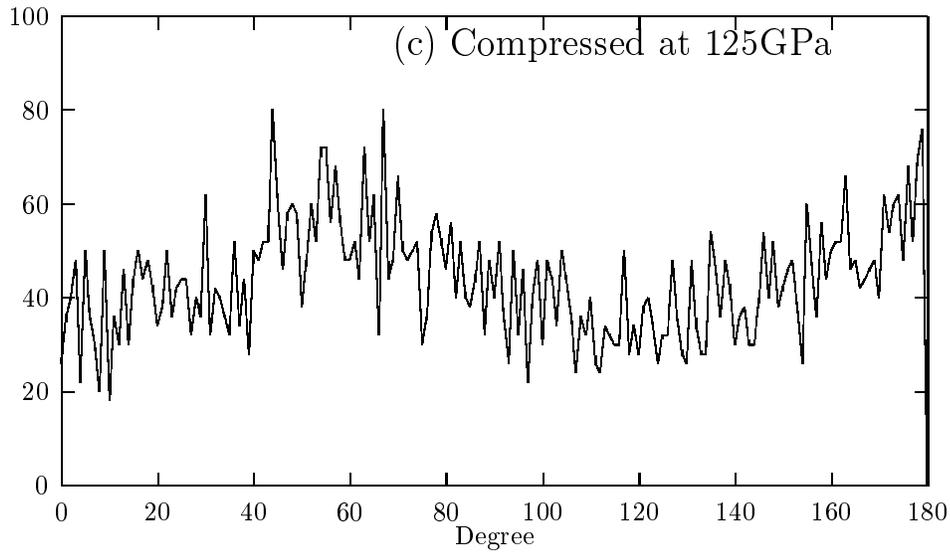
\item Figure 5(c):Distribution of the azimuthal angle 
                  after the compression at 125GPa,
                  in the structure of figure 4(b).

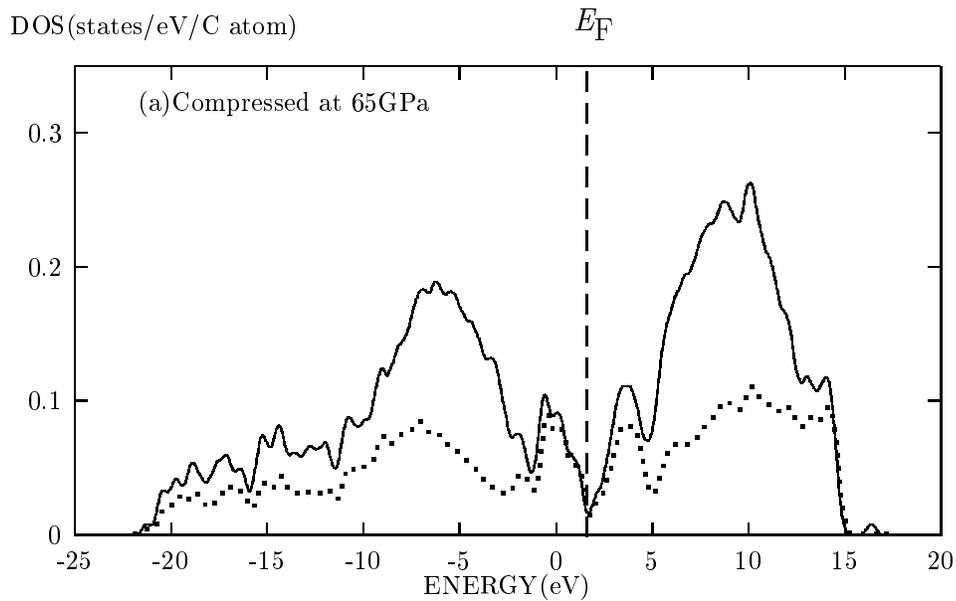
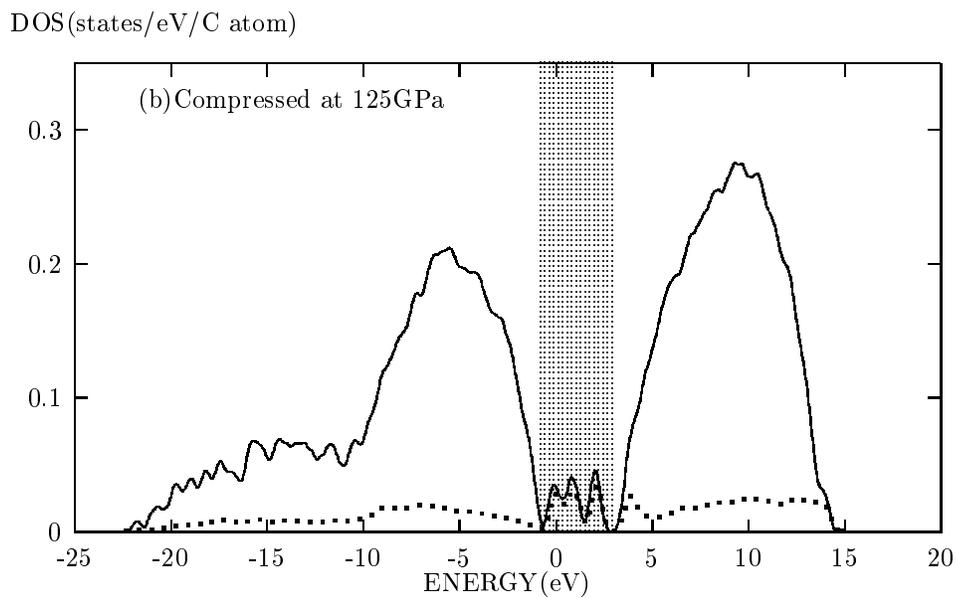
\item Figure 6  :Density of states corresponding to figure 5.
                  The solid line shows the total DOS, 
                  and the dotted line shows the
                  contribution from threefold carbon atoms.  
\item Figure 6(a): DOS after the compression at 65GPa.
\item Figure 6(b): DOS after the compression at 125GPa.

\item Figure 7: An example of  the dynamical effect 
                caused by changes in the pressure through the
                compression is given here.
                Pressure is raised from 0GPa to 65GPa,
                and after that, again raised to 125GPa.
                The early stage of this simulation,
                where the pressure is confined in the range from
                0GPa to 65GPa, corresponds
                to the snapshot given in figure 1.
                These figures show the very slow reconstruction 
                process of the short-range order in the amorphous
                at the high pressure and temperature.

\item Figure 7(a):The change in the applied pressure.

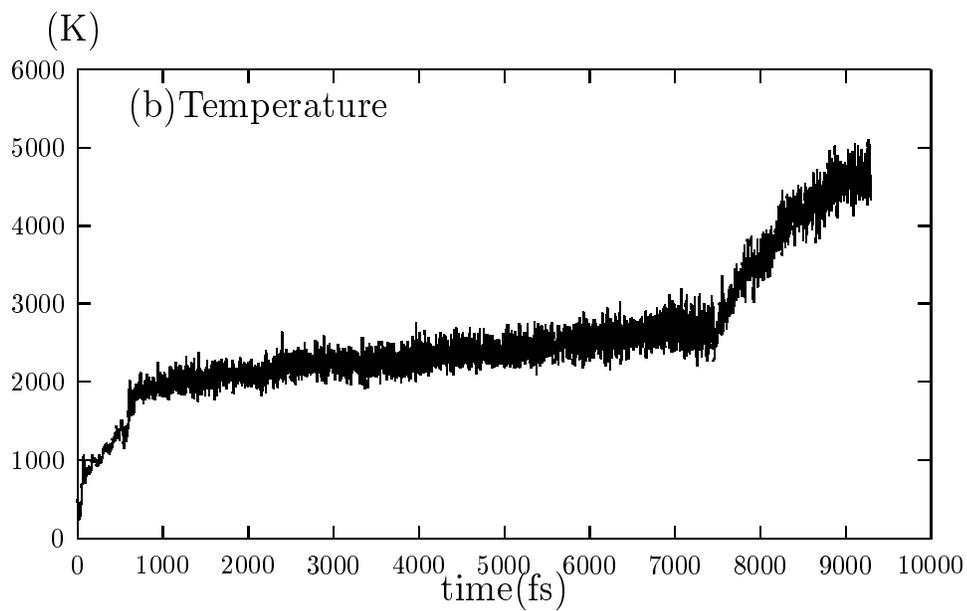
\item Figure 7(b):The variation in the temperature.
                  By raising the pressure again,
                  the temperature increases again.   

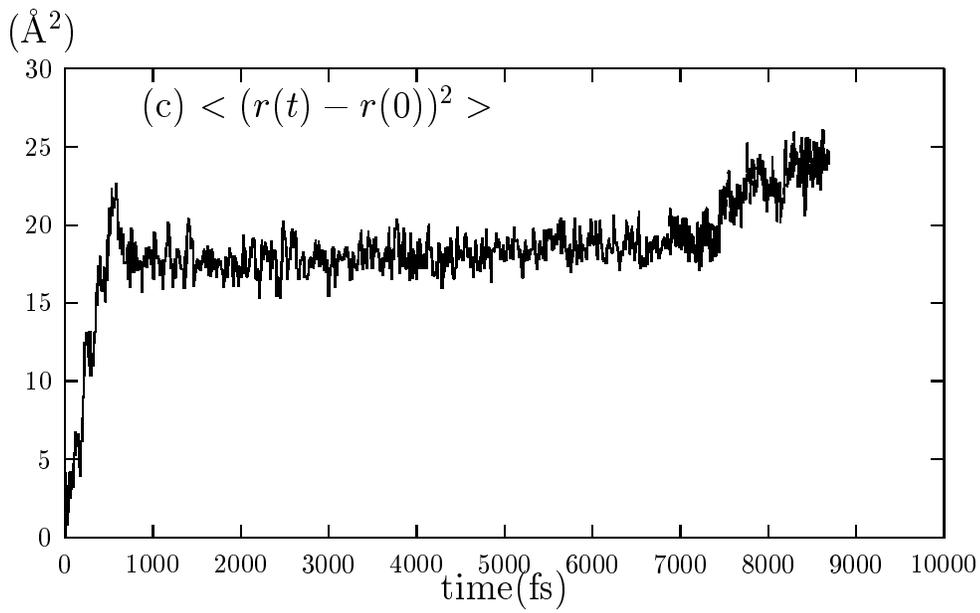
\item Figure 7(c):The mean square of the displacement.
                  By raising the pressure again, the atomic movement
                  becomes more active.

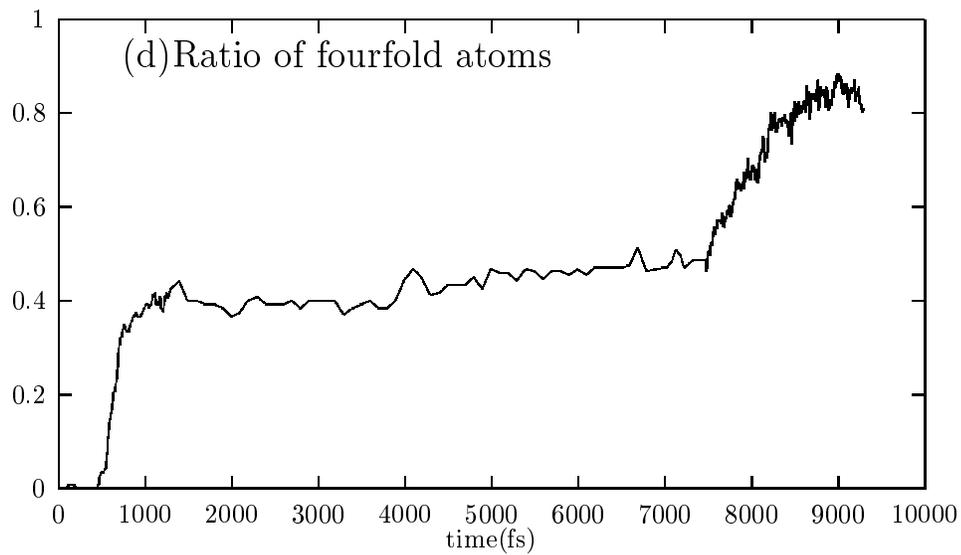
\item Figure 7(d):The ratio of fourfold carbon atoms.
              The increase in this ratio means the formation 
              of the   sp$^3$ bonding.
              By raising the pressure again, 
              much more numbers of the sp$^3$ bonding are generated.

\item Figure 8: Another example of 
                the dynamical effect caused by change in the temperature
                through the compression is given here.
                The pressure is raised from 0GPa to 125GPa.
                In the path, denoted as (A), there is no
                concentration on the temperature.
                In the path (B), the temperature is scaled at 2500K
                after 0.5ps.

\item Figure 8(a):The change in the applied pressure.

\item Figure 8(b):The variation in the temperature. 

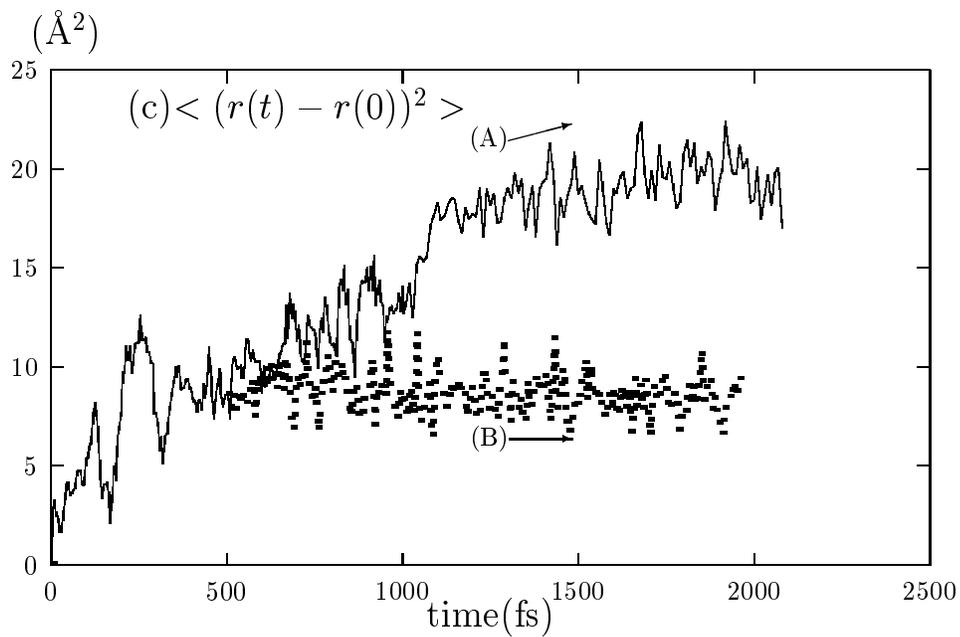
\item Figure 8(c):The mean square of the displacement.
                  In the path (B), owing to the lower temperature,
                  the atomic movement becomes more inactive.

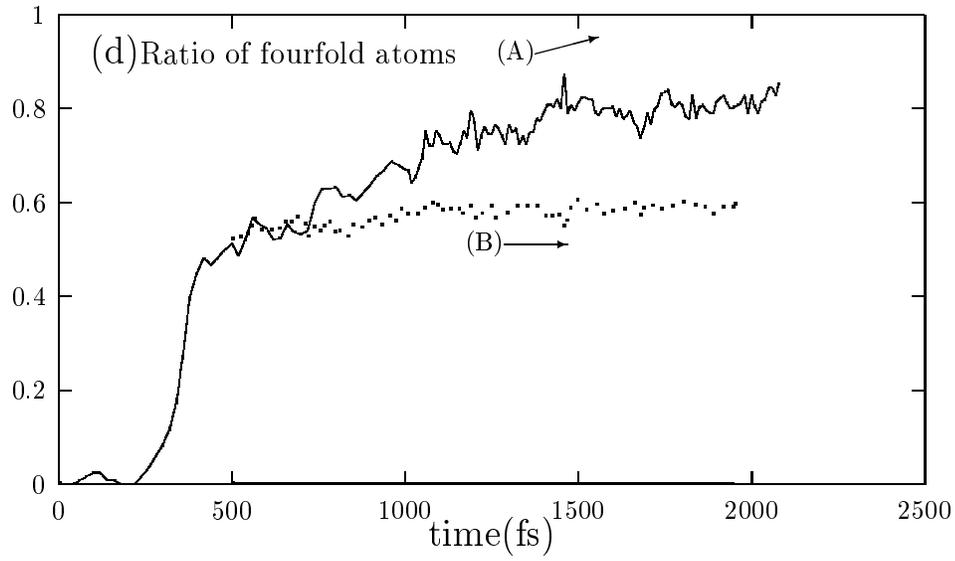
\item Figure 8(d):The ratio of fourfold carbon atoms.
              In the path (B), owing to the lower temperature,
             the transition speed from sp$^2$ to sp$^3$ decreases.

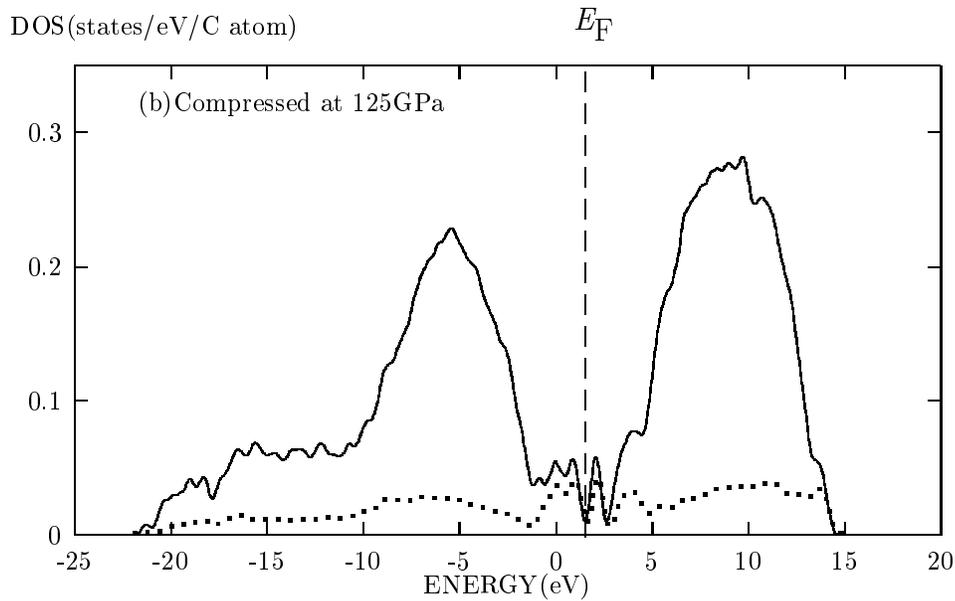
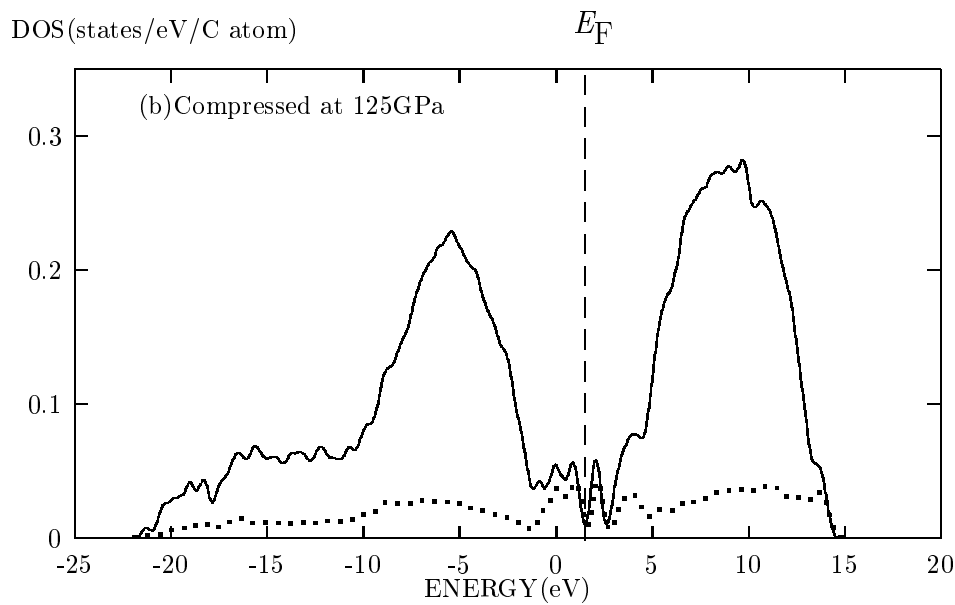
\item Figure 9: The density of states with the presence of hydrogen.
               The solid line shows the total DOS, 
               and the dotted line shows the
               contribution from threefold carbon atoms.
               The cell includes hydrogen atoms at 12.5\% in number.
               Figure 9(a) shows the DOS after the compression
               at the maximum pressure 55GPa. 
               The contribution
               from threefold C atoms 
               is about 40\% in the total DOS.
               Figure 9(b) shows the DOS after the compression
               at the maximum pressure 125GPa.
               The contribution
               from threefold C atoms 
               is about 20\% in the total DOS.

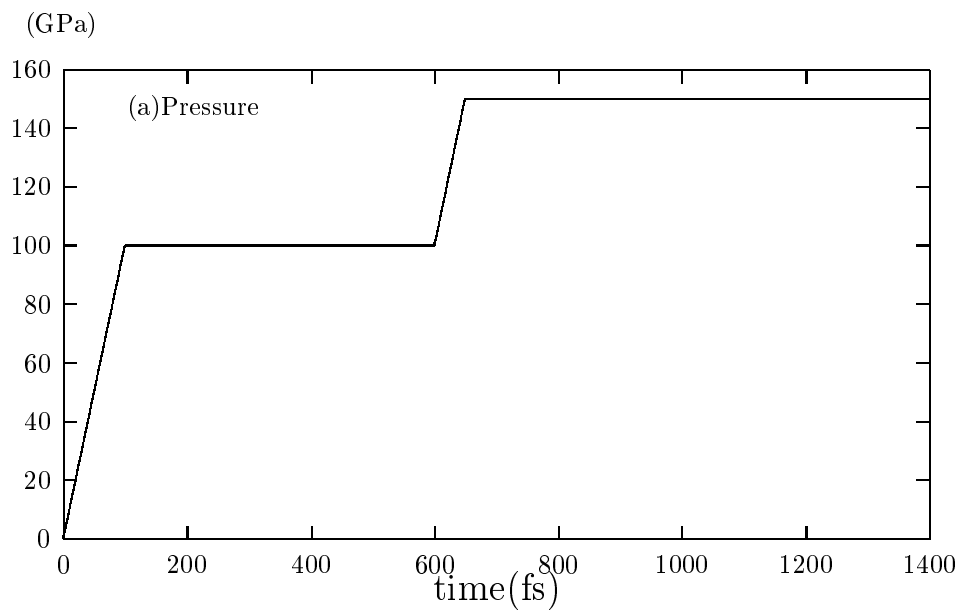
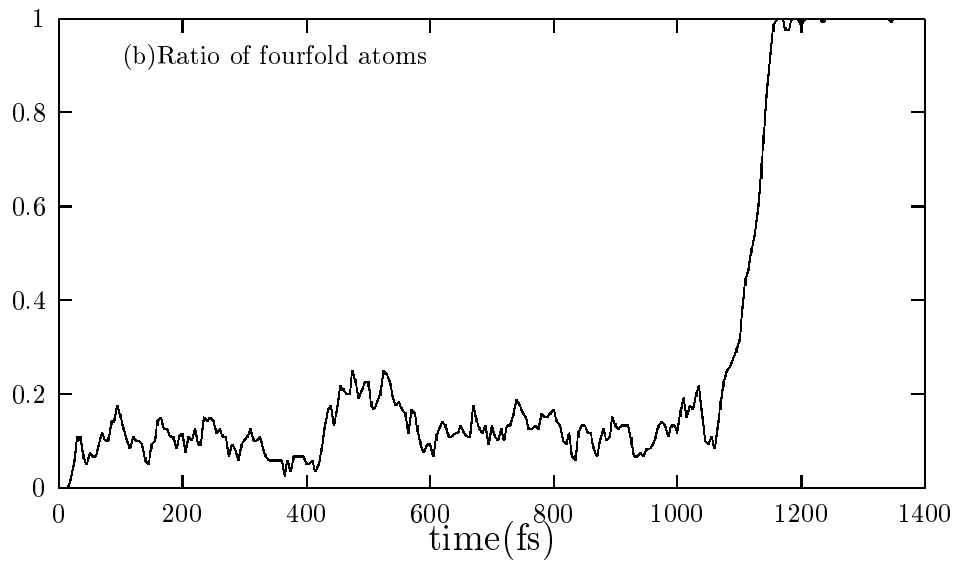
\item Figure  10: This figure shows
                the ratio of fourfold carbon atoms 
                in the pressure-induced structural transition
                from graphite to cubic diamond. 
                The transition from sp$^2$ to sp$^3$ in the whole
                crystal can bee seen in the rapid increase in that ratio,
                changing from 0 to 1. In this simulation,
                the unit cell includes 240 carbon atoms.
                The pressure is raised from 0GPa to 150GPa in initial 0.15ps
                and kept constant. The temperature
                is scaled to be 5000K throughout the simulation.
                The graphite directly turns into the
                stable perfect cubic diamond, 
                taking a reaction path similar to that demonstrated in a
                first principles simulation\cite{G2D}.
\end{itemize}

\begin{thebibliography}{1}
\bibitem{HIRAI} H.Hirai, K.Kondo, N.Yoshizawa, and M.Shiraishi,
                Appl.Phys.Lett.{\bf 64} 1797(1994);
                H.Hirai and K.Kondo, Phys.Rev.B{\bf 51},15555(1995);
\bibitem{GRA_AM}C.S.Yoo and W.H.Nellis,Science {\bf 254}, 1489(1991).
\bibitem{GRA_AM2}T.Sekine, Proc.Jpn.Acad.Ser. B{\bf 68},95(1992).
\bibitem{POLYM} Y.Iwasa {\rm et al.}, Science {\bf 264},1570(1994).
\bibitem{POLYM1} A.M.Rao {\rm et al.}, Science {\bf 259},955(1993).
\bibitem{POLYM2} M.O'Keefe, Nature {\bf 352},674(1991).
\bibitem{PARAM_CH}M.D.Winn,M.Rassinger, and J.Hafner,
                  Phys.Rev.B{\bf 55},5364(1997).
\bibitem{Wentz} R.M.Wentzcovitch, J.L.Martins, and G.D.Price,
               Phys.Rev.Lett.{\bf 70}, 3947(1993);
               R.M.Wentzcovitch, W.W.Schulz, and P.B.Allen,
               {\it ibid.}{\bf 75},3389(1994).
\bibitem{G2D} Y.Tateyama, T.Ogitsu, K.Kusakabe, and S.Tsuneyuki,
              Phys.Rev.B{\bf 54},14994(1996).
\bibitem{DENS_MAT}X.-P.Li, R.W.Nunes, and D.Vanderbilt,
  Phys.Rev.B{\bf 47},10891(1992).
\bibitem{Bundy}F.P.Bundy, J.Chem.Phys.{\bf 38},618(1963),
               {\it ibid.}{\bf 38},631(1963)
\bibitem{DUCLOS} S.J.Duclos, K.Brister, R.C.Haddon,
                  A.R.Kortan, and F.A.Thiel, Nature {\bf 351},380(1991).
\bibitem{KAGAKU}H.Hirai and K.Kondo, Kagaku {\bf 50},368(1995),in Japanese.
\bibitem{C60FCCBAND} S.Saito and A.Oshiyama, Rhys.Rev.Lett.{\bf 66}, 
                    2367(1991).
\bibitem{C60_DIA}H.Hirai,K.Kondo, and T.Ohwada,Carbon {\bf 31},1095(1993).
\bibitem{TEMP_CH} In the compression with the presence of hydrogen,
                  the temperature grows more than that does
                  in the case without hydrogen, since the
                  combination energy between C and H
                  is released together with that between C and C. 
                  However the transition speed itself
                  is not raised so large. 
\bibitem{QUENCH_SPEED} H.Hirai and K.Kondo, Science {\bf 253},772(1991).
\bibitem{COMPRESS_GRA} 
                Even in the pressure-temperature range
                where the C$_{60}$ turns into the amorphous,
                the graphite is stable, while being
                somewhat compressed along c-axis. 
                The simulation, however, shows that
                the compression at a sufficiently high pressure
                and a high temperature 
                distorts the graphite sheets
                and causes the fluctuation
                where the weak inter-layer bridging is being formed.
                If  such a fluctuation can grow large enough,
                the global transition to the diamond will start.
                This accounts for the time-lag between 
                the raise of the pressure and that of the ratio
                of sp$^3$[Fig.10]. Such inter-layer bridging 
                in the fluctuation easily vanishes and does not remain
                after the decrease in the pressure and the temperature.
\bibitem{Xu} C.H.Xu, C.Z.Wang, C.T.Chan, and K.M.Ho, 
J.Phys.Condens.Matter {\bf 4} ,6047(1992).
\end{thebibliography}
\end{document}